\documentclass{iau}
\usepackage{graphicx}

\title[Order and chaos in hydrodynamic BL Her models] 
{Order and chaos in hydrodynamic BL Her models}

\author[Rados\l{}aw Smolec \& Pawe\l{} Moskalik]   
{Rados\l{}aw Smolec
 \and Pawe\l{} Moskalik}

\affiliation{Nicolaus Copernicus Astronomical Centre, \\ ul. Bartycka 18, 00-716 Warszawa, Poland \\ email: {\tt smolec@camk.edu.pl}}

\pubyear{2013}
\volume{301}  
\pagerange{119--126}
\jname{Precision Asteroseismology}
\editors{W. Chaplin, J. Guzik, G. Handler \& A. Pigulski, eds.}
\begin{document}

\maketitle

\begin{abstract}
Many dynamical systems of different complexity, e.g. 1D logistic map, the Lorentz equations, or real phenomena, like turbulent convection, show chaotic behaviour. Despite huge differences, the dynamical scenarios for these systems are strikingly similar: chaotic bands are born through the series of period doubling bifurcations and merge through interior crises. Within chaotic bands periodic windows are born through the tangent bifurcations, preceded by the intermittent behaviour. This is a {\bf universal behaviour} of dynamical systems (\cite[Feigenbaum 1983]{Feigenbaum83}). We demonstrate such behaviour in models of pulsating stars.
\keywords{convection, hydrodynamics, methods: numerical, stars: oscillations, Cepheids}
\end{abstract}


During our study of hydrodynamic BL~Her-type models showing periodic and quasi-periodic modulation of pulsation akin to the Blazhko Effect (\cite[Smolec \& Moskalik 2012]{SM12}) we also found the domains of chaotic behaviour. Here we report the initial analysis of these models focusing on the universal behaviour they display.

All our models were computed with the Warsaw non-linear convective pulsation codes (\cite[Smolec \& Moskalik 2008]{SM08a}). Models have the same mass ($M=0.55{\rm M}_\odot$) and chemical composition ($X/Z=0.76/0.0001$) and were computed along a horizontal stripe ($L=136{\rm L}_\odot$) in the HR diagram with a maximum step in effective temperature of $1$\thinspace K. Convective parameters are the same as in \cite[Smolec \& Moskalik (2012)]{SM12}. We note that the eddy-viscous dissipation was strongly decreased resulting in excessive pulsation amplitudes as compared with observations. Yet, the models show a wealth of dynamical behaviours characteristic for deterministic chaos. Except of period doubling effect (\cite[Smolec et al. 2012]{SmolecEtal12}) such behaviour was not detected in any BL~Her star so far, but these models may help to understand the pulsation of more luminous, longer-period, irregular variables.

\begin{figure}[t]
\begin{center}
\includegraphics[width=5.in]{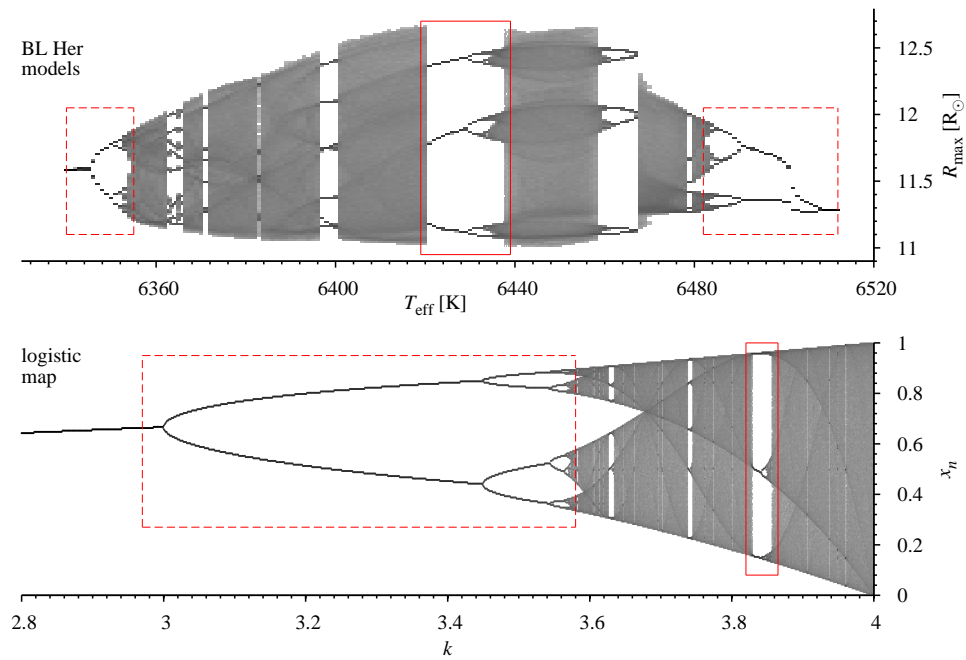} 
\end{center}
\caption{Bifurcation diagrams for hydrodynamic models (top) and the logistic map (bottom).}
\label{fig1}
\end{figure}

In Fig.\,\ref{fig1} (top) we show the bifurcation diagram for our models: the histogram of possible values of maximum radii (gray shaded) plotted as a function of the control parameter for which we choose the effective temperature, $T_{\rm eff}$. This diagram is compared with the bifurcation diagram for the simplest chaotic system, namely the iteration of the logistic map, $x_{n+1}=k x_n (1-x_n)$ (Fig.\,\ref{fig1}, bottom). Despite the huge differences between the two systems striking similarities are apparent.

The chaotic bands are born through the period doubling cascades (within dashed frames in Fig.\,\ref{fig1}). In case of our models the single-periodic (period-one) cycle (one point at the bifurcation diagram) found at both the cool and the hot edge of the computational domain undergoes a series of period doubling bifurcation en route to chaos. Period-two, period-four and period-eight cycles are all detected. In case of the logistic map the same scenario is computed as $k$ is increased beyond $k=3.0$.

Within the chaotic bands a periodic windows of order emerge. There are several such windows in case of our models, e.g. period-5, period-6, period-7 or period-9 windows. We stress that, contrary to recent claims (\cite[Plachy et al. 2013]{Plachy}), stable periodic cycles are not necessarily caused by the resonances among pulsation modes, but may be an intrinsic property of the chaotic systems. We focus our attention on parameter ranges indicated with solid frames in Fig.\,\ref{fig1}. Similarity between our models and the logistic map is apparent again. The well understood properties of simple logistic mapping allow us to explain the dynamics of the much more complex hydrodynamic models. As $k$ ($T_{\rm eff}$) is increased the stable period-3 cycle is born through the tangent bifurcation. The bifurcation is preceded with {\it intermittent} behaviour (\cite[Pomeau \& Mannevile 1980]{PM80}) --  evolution of the system is characterized by long phases of almost periodic behaviour interrupted with shorter bursts of chaos. As $k$ ($T_{\rm eff}$) is increased further, period-3 cycle undergoes a series of period doubling bifurcations to form three chaotic bands. These bands finally merge through {\it interior crises} (\cite[Grebogi, Ott \& Yorke 1982]{GOY82}) -- a bifurcation in which the volume occupied by chaotic attractor suddenly changes. The crises occurs when three chaotic bands collide with the unstable period-3 cycle born along with stable period-3 cycle in the tangent bifurcation. Detailed analysis of the presented models is in preparation. 

{\bf Acknowledgements.} We acknowledge the IAU grants for the conference. This research is supported by the Polish Ministry of Science and Higher Education through Iuventus+ grant (IP2012 036572) awarded to RS.

\end{document}